\documentclass[11pt,twoside]{article}

\usepackage{asp2006}
\usepackage{epsf}
\usepackage{graphicx}
\usepackage{lscape}

\begin{document}
\title{A \textit{Spitzer}/IRS Study of Local Luminous Infrared Galaxies} 
\author{M. Pereira-Santaella\altaffilmark{1}, A. Alonso-Herrero\altaffilmark{1}, G. H. Rieke\altaffilmark{2}, L. Colina\altaffilmark{1}}
\altaffiltext{1}{Instituto de Estructura de la Materia, CSIC, Serrano 121, E-28006, Madrid Spain}
\altaffiltext{2}{Steward Observatory, University of Arizona, 933 North Cherry Avenue, Tucson, AZ 85721, USA}

\begin{abstract}
We present the first results of our program to study a sample of local luminous infrared galaxies (LIRGs, $L_{\rm IR}$ = 10$^{11}$-10$^{12}~L_{\rm \odot}$) with the \textit{Spitzer} infrared spectrograph (IRS). 
In these proceedings we investigate the behavior of the 9.7 \micron\ silicate feature in LIRGs. As opposed to the extreme silicate absorptions observed in ultraluminous infrared galaxies (ULIRGs, $L_{\rm IR}$ = 10$^{12}$-10$^{13}~L_{\rm \odot}$), LIRGs exhibit intermediate silicate absorption features, comparable to those of starburst galaxies.
We also find that most of the LIRGs have the minima of the [NeIII]\slash[NeII] ratio located at their nuclei. It is likely that increased densities in the nuclei are responsible for the smaller nuclear ratios. In the nuclei, it is also possible that the most massive stars are either absent, or still embedded in ultracompact HII regions.
Finally we discuss the possible contribution of an AGN to the nuclear mid-IR emission of the galaxy, which in general is low in these local LIRGs.
\end{abstract}

\section{Introduction and Data Reduction}
Luminous infrared galaxies (LIRGs, $L_{\rm IR}$ = 10$^{11}$-10$^{12 } L_{\rm \odot}$) are an important cosmological class of galaxies as they are the main contributors to the co-moving star formation rate density of the universe at $z$ = 1. Moreover, the mid-IR spectra of high redshift (z $\sim$ 2) very luminous IR galaxies ($L_{\rm IR}$ $>$ 10$^{12} L_{\rm \odot}$) appear to be better reproduced by those of local starbursts and LIRGs.

We obtained \textit{Spitzer}/IRS mapping observations and analyzed archival IRS staring observations of the sample of local (d $<$ 75 Mpc) LIRGs of \citet{AAH06s_mps}.
The maps cover the central 20$\arcsec\times$20$\arcsec$ or 30$\arcsec\times$30$\arcsec$ regions of the galaxies, and use all four IRS modules to cover the full 5-40 $\micron$ spectral range.
The IRS spectral mapping data reduction is described in detail by \citet{MPS2010_mps}. Briefly, we assembled the spectral data cubes from the staring observations with CUBISM. Then we used the low spectral resolution (R $\sim$ 60-120) data cubes to measure the PAH and the 9.7 \micron\ silicate features and we obtained maps of the brightest emission lines by fitting the high resolution (R $\sim$ 600) spectra. Using these emission line maps we calculated ratio maps.
 
\section{The 9.7 \micron\ Silicate Feature}
\begin{figure}[!ht]
\center
\plottwo{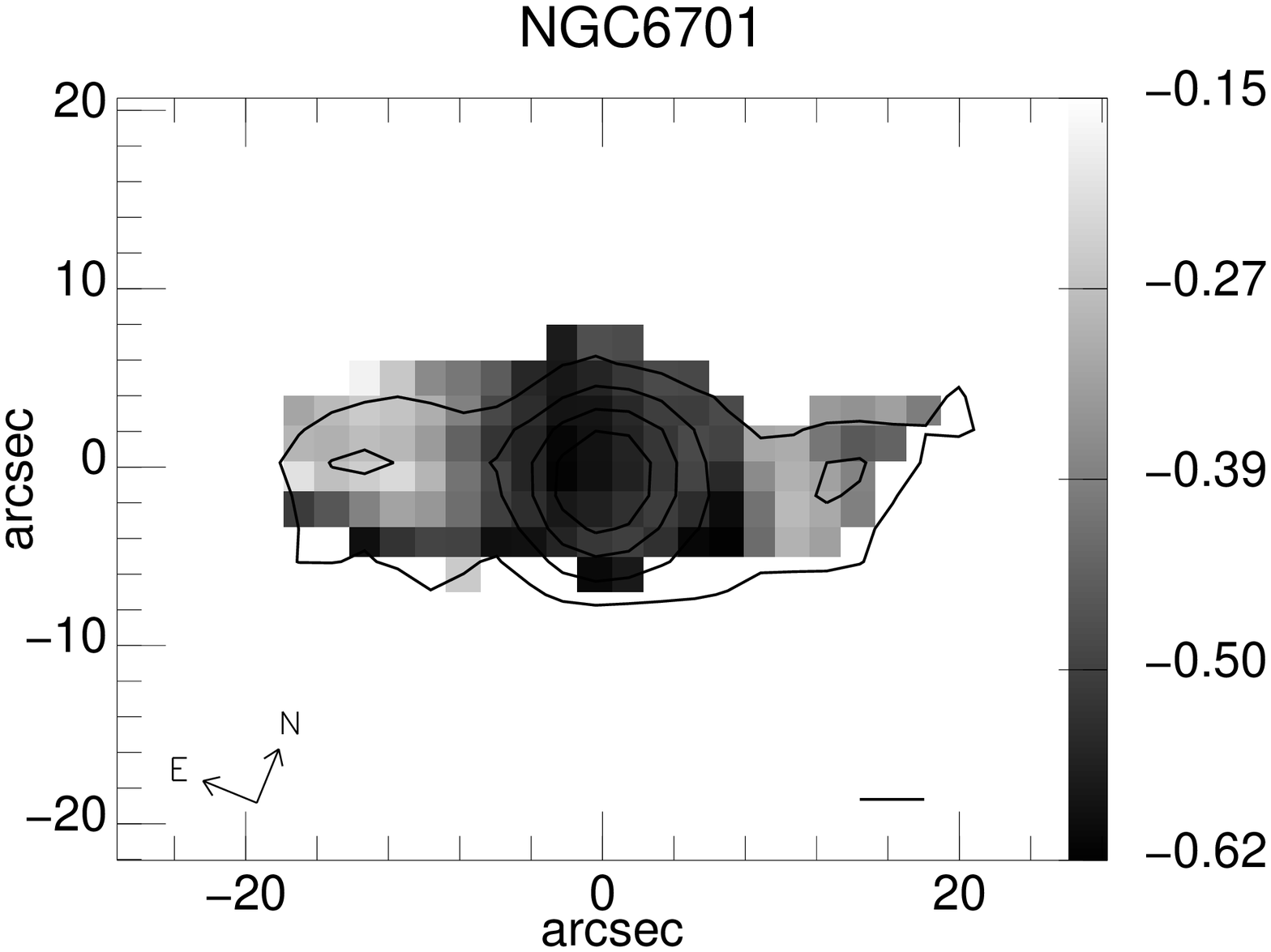}{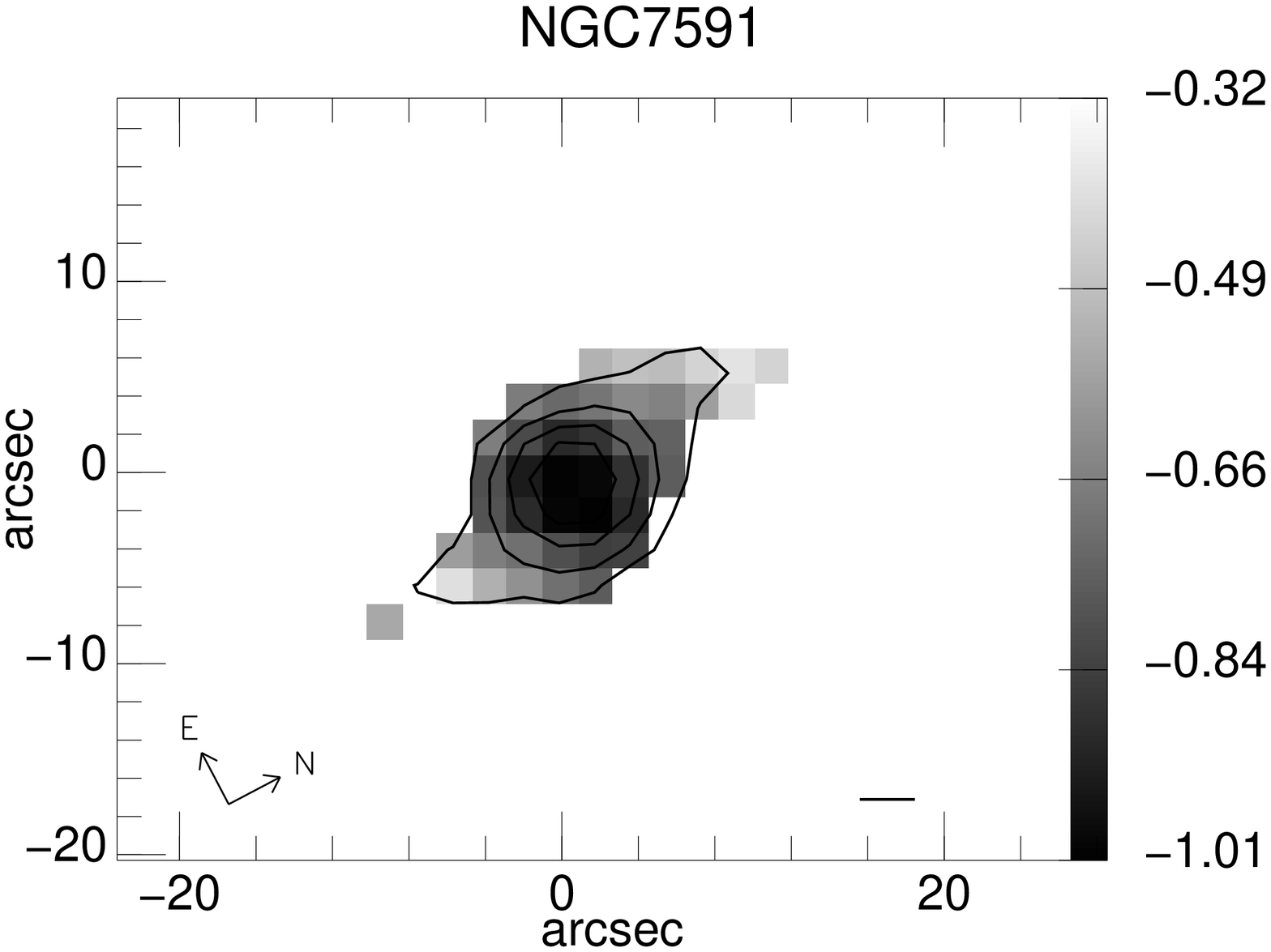}
\caption{Map of the 9.7 \micron\ silicate feature. Dark colors indicate deeper silicate features. The 5.5 \micron\ continuum contours are displayed to guide the eye. The horizontal bar represents 1 kpc.}
\label{fig_mps_silicatos}
\end{figure}

Amorphous silicate grains have a broad spectral feature at 9.7 \micron. Generally this feature is seen in absorption with different strengths. Although in some galaxies, most of them type 1 AGNs, it is seen in emission.
We use the following definition $S_{\rm sil} = ln \frac{f_{\rm obs}(10~\mu m)}{f_{\rm cont}(10~\mu m)}$ to measure the depth of the feature (see \citealt{MPS2010_mps} for details).
We find that the silicate strengths of our sample of LIRGs are relatively shallow compared with those of ULIRGs and comparable to those found in starburst galaxies.
The spatial variations of the silicate strength within a galaxy and from galaxy to galaxy are small (see Figure \ref{fig_mps_silicatos}), although most of the LIRGs have the maxima of the absorption in their nuclei.
The tight range of the silicate strengths among our sample of LIRGs suggest that they may be in part a product of radiative transfer in dusty regions, rather than indicating similar amounts of extinction.

\section{The [NeIII]\slash[NeII] Ratio}
\begin{figure}[!ht]
\plottwo{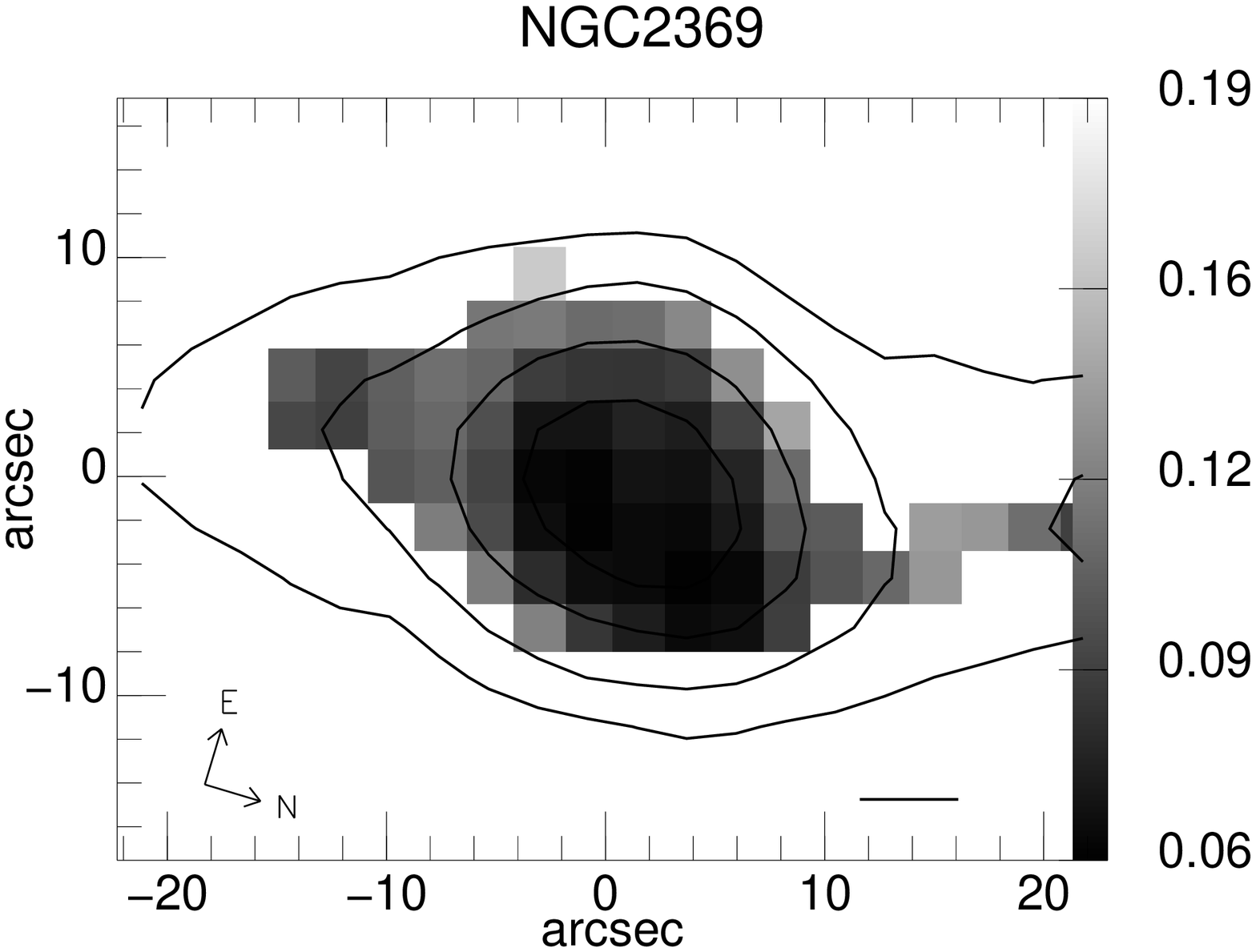}{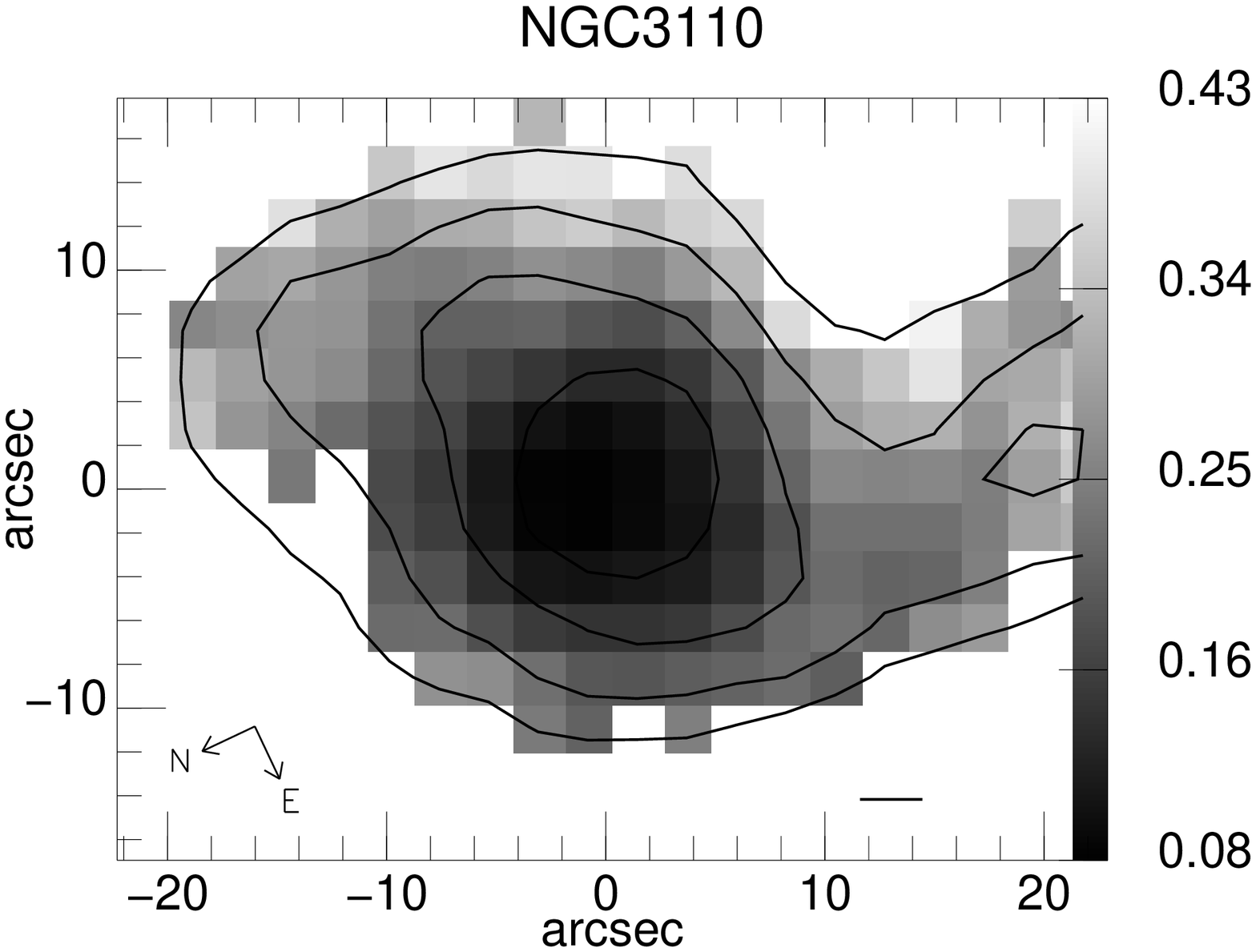}
\caption{IRS SH spectral maps of the [NeIII]\slash [NeII] ratio. The 15 \micron\ continuum contours are displayed to guide the eye.}
\label{fig_mps_cociente}
\end{figure}

The [NeIII] at 15.6 \micron\ and the [NeII] at 12.8 \micron\ emission lines are tracers of recent ($<$ 10 Myr) star formation. The [NeIII]/[NeII] ratio has a strong dependence on the age of the stellar population and metallicity and it falls below 10$^{-3}$ for ages $>$ 6 Myr.
Figure \ref{fig_mps_cociente} shows the [NeIII]/[NeII] ratio maps for two of the LIRGs. Most of the galaxies have the minima of the [NeIII]/[NeII] ratio in their nuclei. The exceptions are those LIRGs classified as active that have more complex morphologies for this ratio. 
The variation of the nuclear [NeIII]/[NeII] ratio is small from galaxy to galaxy and its value is lower than that found in starbursts and below the range predicted by star-formation models. 
Two possible explanations for these nuclear ratios are: (1) the IMF is truncated at $\sim$30 $M_{\rm \odot}$ in the LIRG nuclei; (2) the most massive stars are still buried in ultracompact HII regions.

\section{The AGN Contribution}

AGNs tend to have low PAH equivalent widths (EW), bright high ionization lines (ionization potential $>$ 54 eV) and a hot dust continuum. We used two methods that utilize these characteristics to estimate the AGN contribution in our sample of LIRGs

\begin{figure}[!ht]
\center
\includegraphics[angle=-90, width=0.45\textwidth]{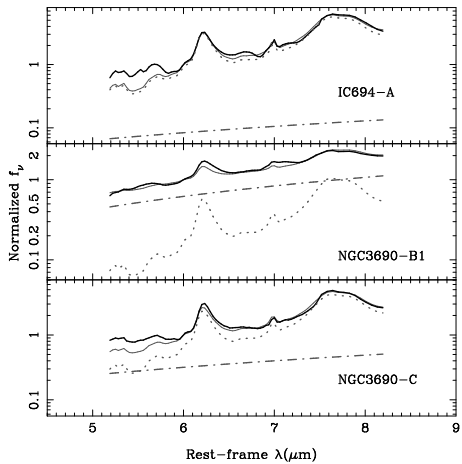}
\caption{Spectral decomposition of the 5-8 \micron\ spectra of the three nuclei of Arp~299. The dot-dash line represents the AGN component (hot dust continuum) whereas the dotted line corresponds to the starburst component (PAH features). A hot dust continuum is clearly detected in NGC~3690-B1  (see \citealt{AAH09_mps}).}
\label{fig_mps_nardini}
\end{figure}

Figure \ref{fig_mps_nardini} illustrates the method proposed by \citet{Nardini2008_mps} to estimate the AGN contribution at 6 \micron\ of three sources of the Arp~299 interacting system. It uses the 5-8 \micron\ spectral region to separate out the starburst component in the form of PAH emission and the AGN component in the form of hot dust continuum.

\begin{figure}[!ht]
\center
\includegraphics[width=0.65\textwidth]{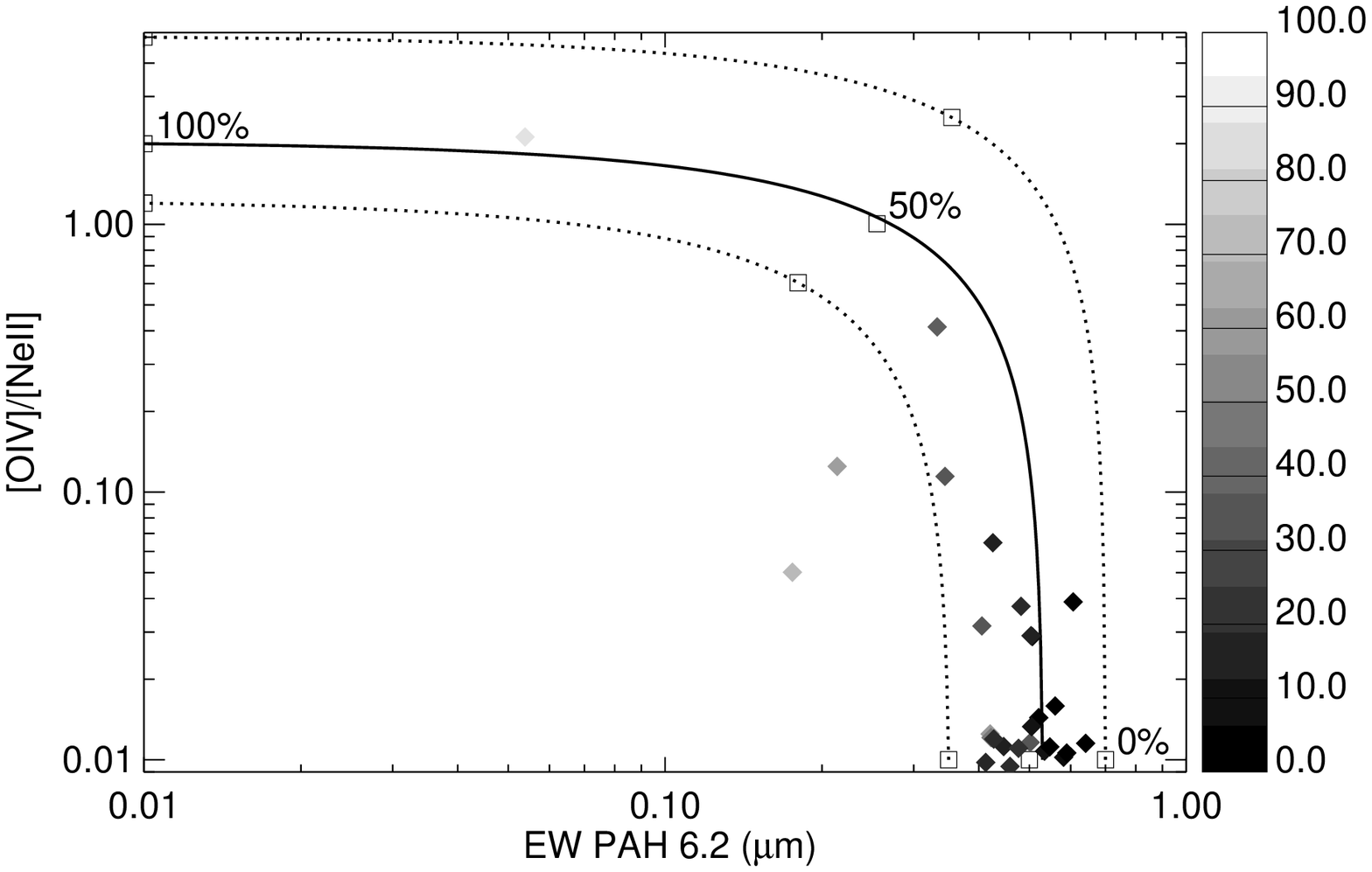}
\caption{[OIV]25.9 \micron\ \slash [NeII] 12.8 \micron\ vs. the EW of the 6.2 \micron\ PAH feature. The diamonds are the IRS\slash staring nuclear values for the \citet{AAH06s_mps} LIRGs.
The solid line is the mixing curve of a pure AGN and a pure starburst and the percentages next to the curve are the AGN contributions. The gray scale on the right side relates the shading of the diamonds to the AGN contribution as calculated from the 5-8 \micron\ spectral decomposition. }
\label{fig_mps_agn}
\end{figure}

Figure \ref{fig_mps_agn} shows the [OIV] 25.9 \micron\slash [NeII] 12.8 \micron\ ratio (55 eV and 22 eV, respectively) versus the equivalent width of the 6.2 \micron\ PAH feature.
The gray scale code corresponds to the estimation from the Nardini et al.~method.
We used the typical [OIV]\slash [NeII] ratios and PAH 6.2 \micron\ EWs of a pure AGN and starburst  for the mixing curve \citep{Brandl06_mps, Dale06_mps, MPS2010AGN_mps}.
Both methods are in good agreement and show that for most of the LIRGs the AGN contribution is small ($<$ 25\%). Only few of these LIRGs (10\% - 20\%) are dominated by an AGN in the mid-IR.

\acknowledgements MP-S acknowledges support from the CSIC under grant JAE-Predoc-2007. Support for this work was provided by NASA through contract 1255094 issued by JPL/California Institute of Technology. 
MP-S, AA-H, and LC acknowledge support from the Spanish Plan Nacional del Espacio under grant ESP2007-65475-C02-01.
AA-H also acknowledges support for this work from the Spanish Ministry of Science and Innovation through Proyecto Intramural Especial under grant number 200850I003.


\end{document}